%% file: ms.tex
\documentclass[12pt,preprint]{aastex}

\usepackage{psfig}

\newcommand{\etal}{\rm et~al.\ } 
\newcommand{\msp}{millisecond pulsar }
\newcommand{\msps}{millisecond pulsars }
\newcommand{\gp}{giant pulse }
\newcommand{\gps}{giant pulses }
\newcommand{\blc}{$B_{\rm LC}$ }
\newcommand{\blcs}{$B_{\rm LC}$}
\newcommand{\pdot}{$\dot{P}$ }

\newcommand{\eav}{\langle E \rangle }
\newcommand{\dmunit}{pc\,cm$^{-3}$}
\newcommand{\dmunits}{pc\,cm$^{-3}$ }
\newcommand{\jyms}{Jy$\rm \mu$s}
\newcommand{\jymss}{Jy$\rm \mu$s }

\newcommand{\simlt}
        {\ifmmode       { \raisebox{-.4em}{$<$}\atop\sim}
           \else        {$\raisebox{-.4em}{$<$}\atop\sim$}
        \fi}

\slugcomment{To be submitted to \apj}

\shorttitle{Giant Pulses from Millisecond Pulsars}
\shortauthors{Knight \etal}

\begin{document}

\title{Green Bank Telescope Studies of Giant Pulses from Millisecond Pulsars}

\author{H. S. Knight
\footnote{Affiliated with the Australia Telescope National Facility,
CSIRO}, M. Bailes}
\affil{Centre for Astrophysics and Supercomputing,
Swinburne University of Technology, P.O. Box 218, Hawthorn VIC 3122,
Australia; hknight@astro.swin.edu.au}
\author{R. N. Manchester}
\affil{Australia Telescope National Facility, CSIRO, P.O. Box 76,
Epping NSW 1710, Australia}
\author{S. M. Ord
\footnote{Current address: RCfTA, School of Physics,
University of Sydney, NSW 2006, Australia}}
\affil{Centre for Astrophysics and Supercomputing, Swinburne University of Technology, P.O. Box 218, Hawthorn VIC 3122, Australia}
\and
\author{B. A. Jacoby
\footnote{Current address: Naval Research Laboratory, Code 7213,
4555 Overlook Avenue, SW, Washington, DC, 20375}}
\affil{Department of Astronomy,  California Institute of
Technology, MS 105-24, Pasadena, CA 91125}

\begin{abstract}
We have conducted a search for giant pulses from four millisecond
pulsars using the 100\,m Green Bank Telescope.  Coherently dedispersed
time-series from PSR~J0218+4232 were found to contain giant pulses of
very short intrinsic duration whose energies follow power-law
statistics.  The giant pulses are in phase with the two minima of the
radio integrated pulse profile but are phase aligned with the peaks of
the X-ray profile.  Historically, individual pulses more than 10-20
times the mean pulse energy have been deemed to be ``giant pulses''.
As only 4 of the 155 pulses had energies greater than 10 times the
mean pulse-energy, we argue the emission mechanism responsible for
giant pulses should instead be defined through: (a) intrinsic
timescales of microsecond or nanosecond duration; (b) power-law energy
statistics; and (c) emission occurring in narrow phase-windows
coincident with the phase windows of non-thermal X-ray emission.  Four
short-duration pulses with giant-pulse characteristics were also
observed from PSR~B1957+20.  As the inferred magnetic fields at the
light cylinders of the \msps that emit giant pulses are all very high,
this parameter has previously been considered to be an indicator of
giant pulse emissivity.  However, the frequency of giant pulse
emission from PSR~B1957+20 is significantly lower than for other
millisecond pulsars that have similar magnetic fields at their light
cylinders.  This suggests that the inferred magnetic field at the
light cylinder is a poor indicator of the rate of emission of giant
pulses.
\end{abstract}


\section{Introduction}\label{sec:introduction}

The Crab radio pulsar was discovered through the direct detection of
strong individual pulses \citep{sr68}.  Further studies revealed that
the strongest pulses followed power-law energy statistics \citep{ag72}
distinct from the Gaussian statistics of the general pulse population
\citep{cor76a}.  In an observation by \citet{lcu+95} around one
in 1200 pulses had an energy greater than 20 times the mean pulse
energy, $\eav$.  Despite this, \citet{cbh+04b} found that at all radio
frequencies phase-coherent summation of the \gps gives a higher
signal-to-noise ratio than summation of all the pulses.
Extraordinarily, the \gps also have structure that is significantly
narrower than the mean pulse.  \citet{hkwe03} observed pulses with
that had structure persisting for less than 2\,ns and inferred that
the brightness temperatures of these pulses are $T_{\rm B}\sim
10^{37}$K.

The young Crab-like pulsar B0540$-$69 in the Large Magellanic Cloud
also emits giant pulses \citep{jr03}.  In 31.2\,hr of observations at
a center frequency of 1390\,MHz \citet{jrmz04} only detected the
integrated emission-profile of PSR~B0540$-$69 at a very low level of
significance.  Despite their difficulty in detecting the integrated
emission, Johnston \etal were able to detect and analyse 141
individual pulses.  The relative ease with which giant pulses can be
seen over large distances has led several authors to advocate their
detection as a way to find extra-galactic pulsars \citep[see
e.g.][]{jr03,cbh+04b}.

To date, no other young pulsars have been found to emit pulses with
the high energies and extremely short durations characteristic of the
giant pulses from the Crab pulsar.  Three young pulsars have been
found to emit narrow pulses of emission showing power-law statistics
\citep{jvkb01,jr02,cjd04}.  However, it is not clear that the pulses
should be classed as true ``giant pulses'' because the power-law tails
have only been seen to extend to low energies.  In addition, the
structure of these events has thus far not been shown to have
timescales as short as those of giant pulses from the Crab pulsar.

The recycled pulsars B1937+21, B1821$-$24, and J1823$-$3021A also emit
giant pulses despite having markedly different periods ($P$) and
period derivatives (\pdot) to the Crab pulsar
\citep{cstt96,rj01,kbmo05}.  One common factor between these \msps,
PSR~B0540$-$69, and the Crab pulsar is that they all have very high
magnetic fields inferred at their light cylinders\footnote{The ATNF
Pulsar Catalogue has been used to obtain the pulsar parameters and
statistics used in this paper.  See:
http://www.atnf.csiro.au/research/pulsar/psrcat.}  $B_{\rm LC} \propto
P^{-2.5}\dot{P}^{0.5}$.  When viewed in the context of the known \msp
population, the three \gp emitters also have very low characteristic
ages $\tau = P/(2\dot{P})$ and very high spin-down luminosities
$\dot{E} \propto P^{-3}\dot{P}$.  PSRs B1821$-$24, B1937+21, and
J0218+4232 have some of the highest X-ray luminosities of all \msps
\citep{bt99,gch+02,cus04,hge+05}.  The emission from all three pulsars
is non-thermal, and the X-ray profiles of PSRs B1821$-$24 and B1937+21
align in phase with their \gp emission \citep{rj01,chk+03}.  Another
field pulsar with a high X-ray luminosity is PSR~B1957+20
\citep{bt99}.  However, no X-ray pulsations have been detected from
this source and it is unclear how much of the emission originates from
the bow-shock between the pulsar wind and the companion wind
\citep{sgk+03}.

In this paper we present the results of a sensitive baseband search
for microsecond-timescale emission from four millisecond pulsars.
Upper limits are placed on emission from PSRs J1843$-$1113 and
J1012+5307, and a new population of short-duration pulses is reported
for PSR B1957+20.  A previously unknown population of giant pulses
from PSR~J0218+4232 is characterized and the results used to clarify
the defining characteristics of giant pulse phenomenology.

\section{Observations and Data Analysis}\label{sec:observations}
All observations were taken using the 100\,m NRAO Green Bank Telescope
(GBT) from 2004 August to 2005 January at frequencies in the ranges of
793-921 and 1341-1469 MHz.  Data were acquired using the Caltech-Green
Bank-Swinburne Recorder II \citep[see][]{jac05}.  This instrument
real-samples one or two dual polarization 64\,MHz-wide bands at the
Nyquist rate.  Software algorithms similar to those described by
\citet{van03} were used to synthesize filter banks.  The first step of
the technique is to Fourier transform the raw voltages to the
frequency domain and divide the spectra into a series of sub-bands.
Each sub-band is multiplied by an inverse-response filter (kernel) for
the interstellar medium (ISM) \citep[see, e.g.][]{hr75,sta98}.  The
sub-bands are then individually Fourier transformed back to the time
domain to give numerous time series, each having coarser time
resolution than the original.  This technique avoids the spectral
leakage suffered by forming the filter bank first and then
transforming back to the Fourier domain to deconvolve.  By splitting
the input signal into sub-bands the dispersive smearing that has
to be accounted for is essentially reduced to that of an individual
sub-band.  This means that to first order the number of samples
required for the initial forward transform is inversely proportional
to the number of sub-bands in the filterbank.  Consequently forming
such a ``coherent filterbank'' uses much shorter transforms than
single-channel coherent dedispersion.  In practical terms this means
the algorithm can use high-speed memory more exclusively and therefore
is computationally faster.

Coherent dedispersion and channel summing were repeatedly applied to
cover a range of dispersion measures (DMs) typically within $\pm
0.1$\,\dmunits of the published pulsar DM.  This guaranteed that our
sensitivity would never be reduced due to DM error.  Data were
square-law detected and combined to give a dataset with bandwidth of
64 or 128 MHz.  These time series were then searched for broad-band
emission by summing adjacent samples at time resolutions between 1 and
128 $\mu$s.  Any two samples with total flux $13\sigma$ ($11\sigma$
for PSR~J0218+4232) or more above the local mean were further reduced
to produce candidate plots for human scrutiny.

Table \ref{tab:observations} summarizes the observations taken.
Columns 1-3 show the pulsar name, center frequency, and bandwidth
respectively.  Columns 4 shows the observation duration, and column 5
shows the number of pulses observed.  The mean pulse energy and
1$\mu$s sensitivity threshold are shown in columns 6 and 7
respectively.  The last column shows the number of individual pulses
detected.  For PSR~J0218+4232 this column shows the number of pulses
detected in each of the ``A'' and ``B'' pulse-phase regions discussed
in Section \ref{sec:properties} and shown in Figure \ref{fig:phen}.
The system equivalent flux densities for the frequency bands centered
at 825-889 and 1373-1437 MHz ranged between 12-14 and 9.1-9.4 Jy
respectively.

\section{Search Results}\label{sec:results}

\subsection{PSR~J0218+4232}\label{sec:0218}

\subsubsection{Properties of the Pulses}\label{sec:properties}

\begin{figure}
  \begin{center}
    \includegraphics[scale=0.6,angle=270]{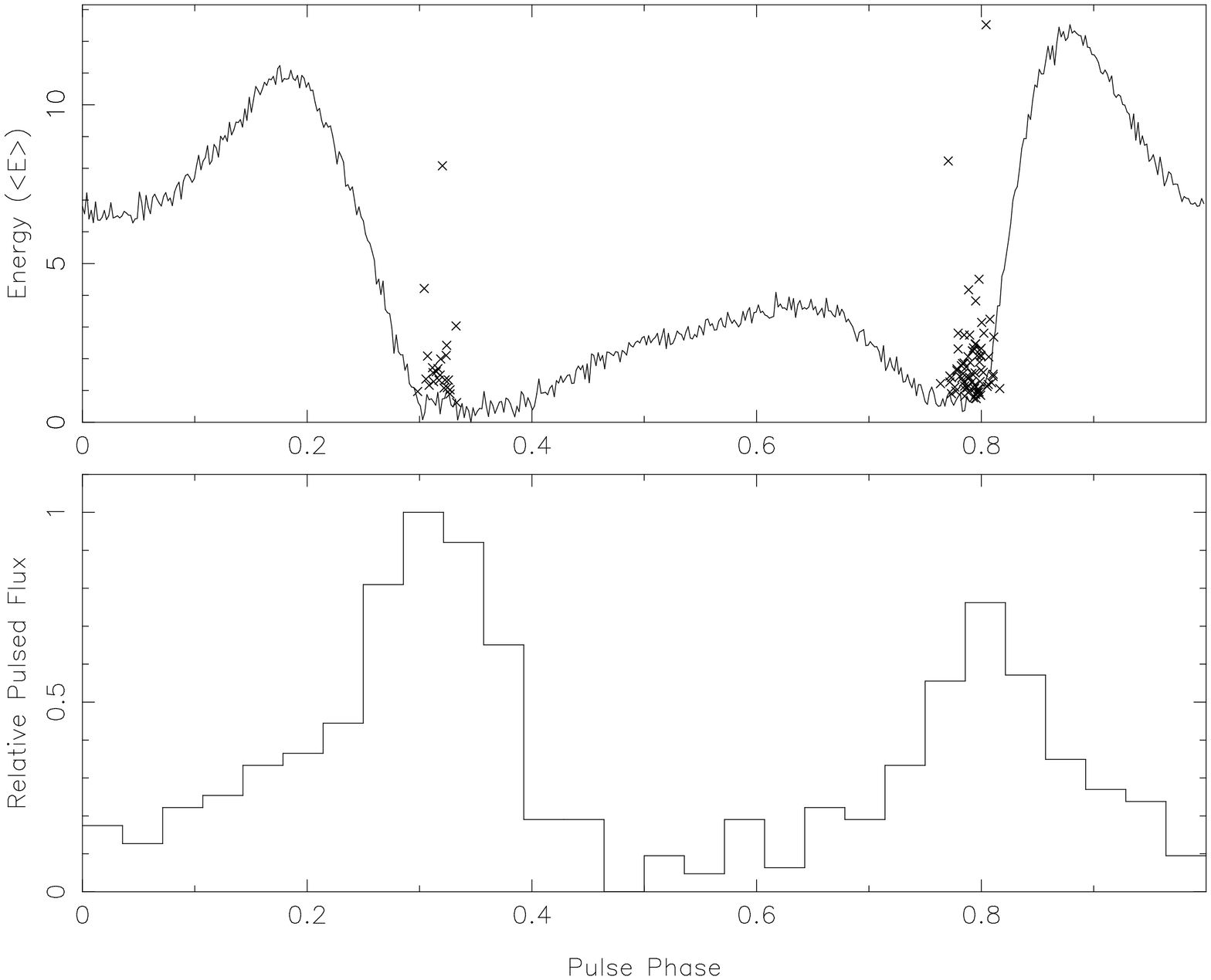}
    \caption{Top: Phases and energies of pulses detected from
    PSR~J0218+4232 in the 2004 August observation are superimposed on
    an integrated pulse profile.  Bottom: The Chandra HRC-S 0.08-10
    keV pulse profile of PSR~J0218+4232 \citep{khs04} has been
    phase-aligned with the radio profile using the absolute timing of
    \citet{rfk+04}.}
    \label{fig:phen}
  \end{center}
\end{figure}

A total of 155 emission events were detected from PSR~J0218+4232.  As
these aligned in two distinct pulse-phase windows (see Figure
\ref{fig:phen}) they are all identified as individual pulses from
PSR~J0218+4232.  Figure \ref{fig:phen} also shows the phases of the
pulses relative to X-ray profile and the full-width, half-maximum
(FWHM) energies of the pulses relative to the mean pulsed flux density.
PSR~J0218+4232 has a significant $\sim$ 50\% unpulsed component
\citep{nbf+95} which was not accounted for in the calculation of the
average pulse energy.  X-ray emission is more prevalent in the earlier
``A'' phase window, but more giant pulses were detected at the later
``B'' phase window.  Our observations therefore show that although
giant pulses in the radio band appear to originate in the same part of
the pulsar magnetosphere as X-ray emission, they are modulated by
different processes.  For the August 857\,MHz observation the ``A''
and ``B'' emission windows spanned 81\,$\mu$s (0.035 periods) and
123\,$\mu$s (0.053 periods) respectively.  Similar widths of 3\% and
4\% of phase were measured for the windows at 1373\,MHz.  The phase
regions in which PSR~B1937+21 emits giant pulses are much narrower.
At 1650\,MHz its two windows are 10.7 and 8.2\,$\mu$s wide, or 0.007
and 0.005 periods wide \citep{spb+04}.  The giant pulses found on the
main emission component of PSR~J1823$-$3021A at 685\,MHz have a
similar phase range to the pulses of PSR~J0218+4232 of about 0.04
periods or 220\,$\mu$s \citep{kbmo05}.

\begin{figure}
  \begin{center}
    \includegraphics[scale=0.6,angle=270]{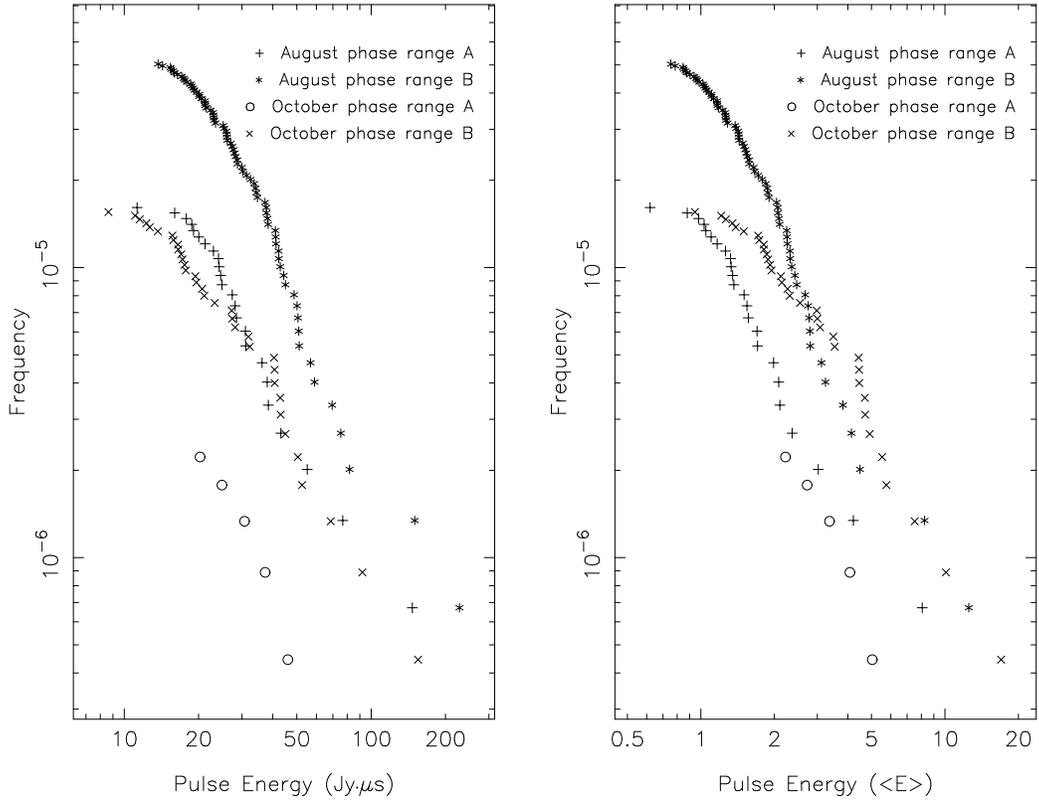}
    \caption{ Cumulative distribution of giant pulse energies for
    observations of PSR~J0218+4232 centered at 857\,MHz when viewed in
    terms of absolute energy (left panel) and relative to the mean
    pulse energy (right panel).}  
    \label{fig:cumu}
   \end{center}
\end{figure}

The Crab pulsar and PSR~B1937+21 emit 10-20$\eav$ pulses at high
rates, and so energy thresholds in this range have been used to
distinguish giant pulses from ordinary emission \citep{ag72,cstt96}.
Only 3 of the 139 pulses seen at 857\,MHz from PSR~J0218+4232 had
energies greater than $10\eav$, and so this pulse population is not
particularly strong compared to the giant pulse populations of the
Crab pulsar and PSR~B1937+21.  However, the argument that these pulses
arise from the same giant pulse emission mechanism is compelling.
Firstly, the cumulative distribution of pulse energies shown in Figure
\ref{fig:cumu} shows that the strongest pulses have power-law
statistics.  The tapering off at low energies is due to the widths of
the pulses being underestimated due to noise.  The pulses are very
narrow and align in phase with the non-thermal X-ray pulses.  All
these properties are shared by the giant pulses of the Crab pulsar and
PSR~B1937+21.  In addition, the pulses from PSR~J0218+4232 occur at
the minima of the integrated emission profile and therefore do not
contribute to the main emission components.  Consequently they cannot
be interpreted as strong ``ordinary'' pulses.  The pulses from
PSR~J0218+4232 therefore demonstrate that the giant pulse phenomenon
can no longer be defined through arbitrary bounds on pulse energy.
Better phenomenological criteria are narrow pulse-widths, power-law
statistics, and emission occuring in narrow phase-windows that align
with non-thermal X-ray emission.

\citet{jr04a} also suggested that power-law statistics and emission at
special phases were the defining characteristics of giant pulses.
Their filter bank observations in previous work \citep{rj01} were
unable to constrain the width of the giant pulses from PSR~B1821$-$24.
Our observations have shown that PSRs J1823$-$3021A
\citep[see][]{kbmo05} and J0218+4232 have intrinsically narrow pulses.
We argue that giant pulses always have narrow widths, and that this
property can be added to those presented by Johnston \& Romani in
defining the giant pulse phenomenon.

The fraction of pulses detected at phase ``A'' almost halved
from 0.24 in August to 0.13 in October.  As the search did not
discriminate on the basis of phase, this difference could be
interpreted as being due to variation in the rate of giant pulse
emission for each phase window.  However, the rate change only becomes
readily apparent when viewed in terms of the detection counts
regardless of pulsar flux (see left panel of \ref{fig:cumu}), and not
when viewed in terms of energy relative to the mean pulse-energy (see
right panel of \ref{fig:cumu}).  Small number statistics are therefore
a more likely cause of the disparity- the non-detection of $\sim 6$
low-energy pulses can explain the rate change.

\subsubsection{Comparison of Emission Rates}\label{sec:rates}

The probability of a pulse having energy greater than $E_{0}$ can be
expressed as:

\begin{equation}
P(E > E_{0}) = K E_{0}^{-\alpha}.
\end{equation}

\noindent Here $E_{0}$ is in units of the mean pulse energy.
Integrating gives an expression for the fraction of pulse flux emitted
in the form of giant pulses of energies greater than $E_{0}$:

\begin{equation}
S_{\rm GP}(E > E_{0}) = \frac{K \alpha}{\alpha-1} E_{0}^{1-\alpha}.
\end{equation}

The best fits for the 857\,MHz pulses with energies greater than
25\,\jymss are shown in Table \ref{tab:powerlawfits}.  No satisfactory
fit was obtained for the October ``A'' pulses.  Estimates of the
relative rate at 1373\,MHz and rates for other pulsars are also shown.
The first three columns show the pulsar, center frequency, and
the phase range the power law is valid for.  Columns 4 and 5 show the
best fits for $K$ and $\alpha$.  The probability that a pulse has $E >
20\eav$ is shown in column 6.  Columns 7 and 8 show the fraction of
flux that is emitted as giant pulses of energies greater than $20\eav$
and $0.1\eav$ respectively.

The power-law energy distributions of the Crab pulsar, PSR~B0540$-$69,
and PSR~B1937+21 (at 430\,MHz) do not extend to energies as low as
0.1$\eav$.  \citet{spb+04} find that at 1650\,MHz the giant pulses
from PSR~B1937+21 extend to energies of 0.016-0.032 $\eav$.  The
power-law exponents for the millisecond pulsars PSR~B1821$-$24 and
PSR~J1823$-$3021A are poorly known.  However, at 1400-1500\,MHz they
emit a giant pulse of more than $28\eav$ at frequencies of $\sim 8.5
\times 10^{-7}$ and $\sim 4.6 \times 10^{-6}$ respectively
\citep{rj01,kbmo05}.  For comparison, the work of Soglasnov \etal
gives for PSR~B1937+21 an emission rate of $P(E > 28\eav) = 2.6 \times
10^{-6}$.  The observed pulse energy distribution is the product of
the intrinsic distribution and the spectra of propagation effects such
as interstellar scintillation.  Scintillation is particularly strong for
PSR~B1937+21 on timescales of minutes at frequencies in the vicinity
of 1-2\,GHz and could potentially lead to different studies obtaining
different results.  \citet{kt00} obtain parameters for PSR~B1937+21 at
1420\,MHz of $\alpha = 1.8$ and $P(E > 28\eav) \sim 4.0 \times
10^{-7}$ which are quite different from those found by Soglasnov \rm et
al.\

It is apparent in Table \ref{tab:powerlawfits} that PSR~J0218+4232 has
a much lower rate of giant pulse emission than other giant pulse
emitters.  The total fraction of its pulsed energy emitted in the form
of \gps with energies greater than $0.1 \eav$ is about 0.1\%.  Such
giant pulses can occur at rates of up to one per $\sim 200$ pulsar
rotations.  If the cut-off point of the power law occurs at
$0.1\eav$ then the $\sim 10$\% of the pulse profile where the giants
occur should have a flux enhancement caused by the giant pulses of
$\sim 1$\% of the mean flux density.  If the power law extends to
$0.01\eav$ then the flux enhancement increases to the 10\% level.
Extension to energies much lower than $0.01\eav$ does not seem
plausible given the lack of large components in the emission regions
of the giant pulses.

The power-law fit for the ten most energetic pulses seen at 1373\,MHz
is summarized in Table \ref{tab:powerlawfits}.  A $20 \eav$ pulse at
1373\,MHz is emitted about 1.7 times more frequently than the August
857\,MHz ``B'' pulses.  It should be noted that the formal uncertainty
on $\alpha$ of $\pm 0.1$ makes this estimate somewhat uncertain.

\subsubsection{Pulse Durations}\label{sec:durations}

\begin{figure}
  \begin{center} \includegraphics[scale=0.6,angle=270]{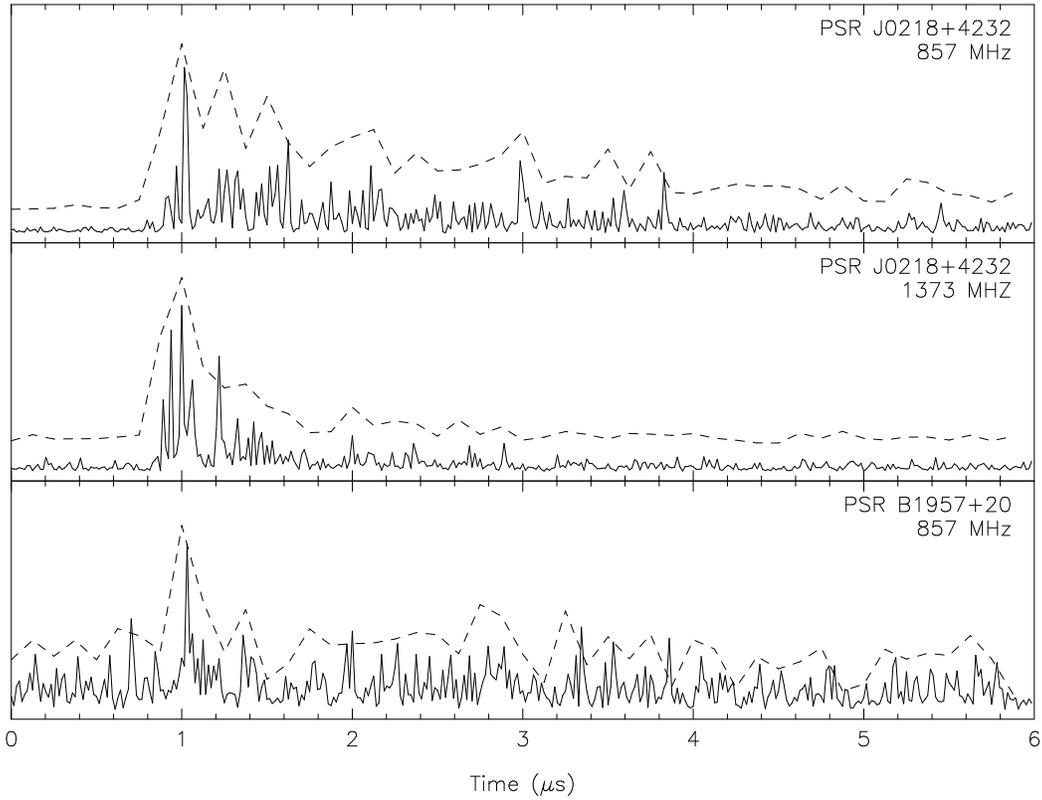}
    \caption{ The intensity of the central portion of the strongest
    giant pulses from PSR~J0218+4232 and PSR~B1957+20 when seen with
    a sampling interval of 15.625\,ns (solid line) and 125\,ns (raised dotted line).}
    \label{fig:best} \end{center}
\end{figure}

All the 857\,MHz pulses had FWHM durations of $\le 3.2$\,$\mu$s.  The
strongest pulse had a FWHM duration of 2.6\,$\mu$s.  To investigate
the possibility of substructure, this pulse was coherently dedispersed
at a time resolution of 15.625\,ns.  The initial portion of the pulse
is shown in the top panel of Figure \ref{fig:best}.  The finite rise
time is only resolved at sampling intervals less than 125\,ns and
persists if the DM is slightly altered.  At high time resolution the
noise statistics are better modelled using chi-squared distributions
than with the standard Gaussian approximation.  The noise statistics
therefore become more positively skewed at higher time resolutions and
so much of the substructure seen at 15.625\,ns time resolution is
likely to be spurious.

The strongest pulse seen at 1373\,MHz as shown in the middle panel of
Figure \ref{fig:best} is significantly narrower.  At a time resolution
of 125\,ns it is of order 500\,ns wide.

Strong spikes following the main emission peak persist for about 8
times longer at 857\,MHz than at 1373\,MHz, and so the pulse widths
are roughly consistent with the $\nu^{-4.4}$ scaling law of
Kolmogorov-spectrum interstellar scattering \citep{bcc03}.  We think
the finite rise-time seen at both 857 and 1373\,MHz is a
consequence of propagation through a thick scattering screen
\citep[see, e.g.][]{wil73} rather than intrinsic substructure.

\subsubsection{Timing of Giant Pulses}\label{sec:timing}

The giant pulse emission of PSR~J0218+4232 occurs over much narrower
ranges of pulse phase than the integrated pulses.  It is therefore important
to consider whether timing of PSR~J0218+4232 can be improved by
timing only the giant pulses.  We formed a standard profile from the
brightest giant pulse and cross correlated the giant pulses with it to
obtain an arrival time.  The 56 giant pulses in the August observation
in phase range ``B'' that had arrival time errors less than
0.5\,$\mu$s had an rms residual of 24\,$\mu$s.  The error in arrival
time for the whole group was therefore about 3\,$\mu$s.  However our
timing of the mean profile for this observation obtains an rms
residual of 6\,$\mu$s using 16.8\,s integrations, which we would
expect to improve significantly with increased integration.  Therefore
conventional timing gives superior results to timing using giant
pulses.

\subsection{PSR~J1012+5307}\label{sec:1012}

PSR~J1012+5307 is a 5.3\,ms pulsar with a characteristic age of
8.6\,Gyr \citep{lcw+01}.  It has a \blc 68 times smaller than that of
PSR~B1937+21.  No individual pulses were detected from PSR~J1012+5307.
Our result is consistent with the fact that to date no \msps with low
values of \blc and large characteristic ages have been observed to
emit giant pulses.  \citet{es03a} observed PSR~J1012+5307 for 1800\,s
at 1380\,MHz using the Westerbork Synthesis Radio Telescope.  With a
sampling interval of 51.2\,$\mu$s they detected 70 individual pulses
with energies of up to five times the mean pulse energy.  Our
observations establish that it is very unlikely that the pulses
uncovered by Edwards \& Stappers have the short $\simlt 1$\,$\mu$s
timescales characteristic of the giant pulses of PSR~B1937+21.

\subsection{PSR~B1957+20}\label{sec:1957}

PSR~B1957+20 has the third highest \blc of all \msps and was therefore
targeted by \citet{kbmo05} as a potential source of \gp emission.
Knight \etal failed to detect any pulses in 7700\,s of observations at
a center frequency of 685\,MHz using the Parkes Radio Telescope.  It
is well known that the pulsar wind of PSR~B1957+20 causes gas to be
ablated from its companion \citep{fbb+90,ks90}.  This ionized gas
causes eclipses at orbital phases ($\phi$) near 0.25.  Knight \etal had
suggested that the gas could scatter-broaden any giant pulses beyond
reasonable detection levels.  However, significant broadening cannot
occur at all orbital phases, as in 8003\,s of observations using the
GBT we detected four narrow pulses from PSR~B1957+20.  Our
observations spanned $0.40 < \phi < 0.67$; the earliest pulse arrived
at $\phi=0.41$.

To estimate the energies of the pulses we formed a 512-bin profile of
each pulse and calculated the FWHM energy.  The pulses had energies of
4.5-8.6\,$\eav$.  At this coarse time resolution virtually all of the
pulse flux for these pulses is encompassed in our estimate.
Adjustment of the DM used for coherent dedispersion and channel
summing causes changes in the noise characteristics of the on-pulse
region.  Peak intensity, pulse morphology, and pulse width all vary
with DM, and so determination of the true DM and therefore true pulse
shape becomes dependent on the exact criteria used to optimize DM.
The bottom panel of Figure \ref{fig:best} shows the strongest pulse.
Although the main pulse component appears very narrow, there is a very
weak underlying emission region about it of microsecond duration.
Other pulses optimize at DMs that differ by $O(10^{-3})$\,\dmunit.  We
think that this DM uncertainty is caused by the low signal strength of
the pulses.  It means that the profiles shown at high time resolution
do not necessarily represent the true pulse form.  For weak pulses
like these, we suggest that the true nature of the pulses in terms of
their individual DMs, intrinsic widths, and substructure requires the
DM to be accurately determined via multi-frequency observations.

\begin{figure}
  \begin{center} \includegraphics[scale=0.6,angle=270]{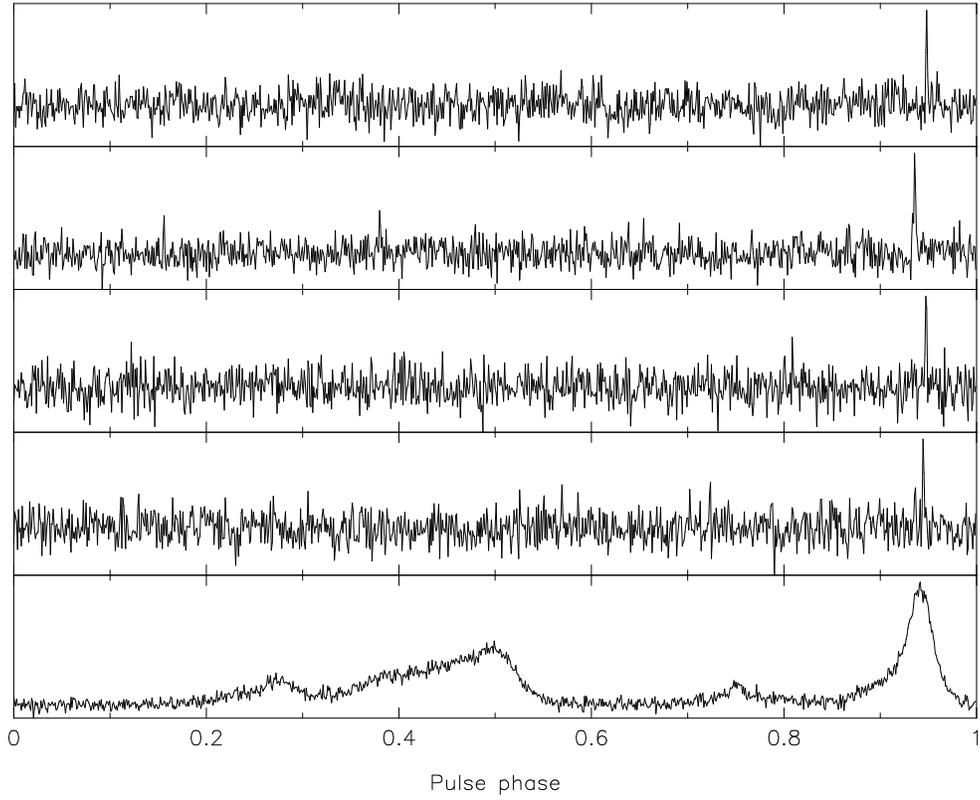}
    \caption{Relative phases of the individual pulses detected from
    PSR~B1957+20.  The top four panels show the intensities of the
    pulses and the bottom panel shows the intensity of the integrated
    pulse profile.}
    \label{fig:bw_phases}
  \end{center}
\end{figure}

The four pulses fall in a narrow 20\,$\mu$s pulse window that covers
the peak of the main emission component (see Figure
\ref{fig:bw_phases}).  The other pulse components of the integrated
profile are 50\% or more weaker than the main component.  It is
reasonable to suppose that PSR~B1957+20 might emit narrow pulses
similar to those observed that are phase-aligned with the other
components.  If these pulses exist and are amplitude modulated in a
similar fashion to the ``ordinary'' pulse emission, they would be 50\%
or more weaker than the main-component pulses we see.  As our initial
detection threshold was 13$\sigma$ and the main-component pulses were
detected at 14-16$\sigma$, our observations do not place good bounds
on the existence of such pulses.

\subsection{PSR~J1843$-$1113}\label{sec:1843}

PSR~J1843$-$1113 is a solitary 1.8\,ms pulsar with a characteristic
age of $\sim 3$Gyr.  Its \blc is very high --- about 0.2 times that of
PSR~B1937+21.  No spikes of broad-band emission were detected from it,
suggesting that if it does emit giant pulses they are very weak and/or
infrequent.  Because PSR~J1843$-$1113 is close to the plane of the
Galaxy and has a relatively high DM we cannot rule out that it emits
pulses similar to those of PSR~B1957+20, but which are scatter-broadened
beyond our sensitivity limits.

\section{Discussion}\label{sec:discussion}

\subsection{Pulse Populations}\label{sec:populations}

\citet{jkl+04a} reported the detection of an unresolved $\sim 129
\eav$ (925\,\jyms) large-amplitude pulse from PSR~B1957+20 in
observations centered at 610\,MHz.  As this pulse is much larger than
any detected in our observations it is instructive to consider whether
it constitutes: (a) a member of a different pulse population of longer
intrinsic duration; (b) the very high-energy tail of the distribution
we observed; or (c) some sort of noise event that is unrelated to the
pulsar.  If the pulse reported by Joshi \etal is as broad as their
sampling time (258\,$\mu$s) then our 825\,MHz detection threshold for
summing two 128\,$\mu$s samples of $45 \eav$ should have found similar
pulses, and so our observations are not consistent with hypothesis
(a).  Alternatively if we assume the pulses follow a $\alpha = -1.4$
energy distribution then our observations imply that a $129 \eav$
pulse should be emitted on average once every 61\,hr.  Given that
Joshi \etal observed PSR~B1957+20 for $\le 1$\,hr and much steeper
power-law distributions have been found for other giant pulse emitters
\citep[see, e.g.][]{rj01} we find hypothesis (b) untenable.  The
detection criterion used by Joshi \etal was that a pulse must exceed
$3.5\sigma$ in two bands.  With an rms of $\sim$ 1 Jy over their
16\,MHz band it is apparent that a bare detection corresponds to
1300\,\jyms, or $180 \eav$.  For a normally distributed noise floor
approximately eleven noise spikes would be expected to exceed this
threshold in their sample of $\sim 10^{6}$ pulses.  The fact that
multiple noise spikes with higher energies than the pulse are expected
to be present in the Joshi \etal data makes it difficult to argue that
the pulse is not background noise.  This in turn implies that we have
presented the first evidence for a population of giant pulses
from PSR~B1957+20.

Joshi \etal also reported the detection at 610\,MHz of three
unresolved large-amplitude $\sim 258$\,$\mu$s wide pulses from
PSR~J0218+4232 with energies of 48-51\,$\eav$.  The event rate of this
pulse population is $P(E>48\eav) = 1.4 \times 10^{-6}$, which is 40
times higher than our August rate for the ``B'' phase range of $P(E >
48\eav) = 3.6 \times 10^{-8}$.  Our August detection threshold for
summing two 128\,$\mu$s samples of $7 \eav$ means that we should have
easily detected the Joshi \etal pulses.  Therefore the pulses of Joshi
\etal are not a separate pulse population that is simply stronger than
the one we observed.  Furthermore, the Joshi \etal pulses occur at a
different phases to the pulses we saw, so the hypothesis that Joshi
\etal were extremely fortunate in detecting the high-energy tail of
our population is not at all plausible.  Pulses similar to those
reported by Joshi \etal should also have been seen by \citet{es03a},
who did not detect any pulses above $26 \eav$ in an 1800\,s
observation centered at 328\,MHz.  If the noise floor of Joshi \etal
is normally distributed then approximately 36 noise spikes in their
sample would be expected to exceed their criterion of $3.5 \sigma$ in
both bands and therefore have a similar energy to the pulses reported.
The three pulses are therefore not distinguishable from background
noise and are likely to be spurious.  The only type of strong pulses
not ruled out by our data reduction are those with timescales
comparable to PSR~J0218+4232's 2.3\,ms pulse period.  However, such
pulses probably would have had substructure detectable in our
searches.  It is more likely PSR~J0218+4232 only emits one population
of strong pulses and that is the population of giant pulses unveiled
by our observations.

\subsection{Giant Pulses from PSR~B1957+20}\label{sec:1957gp}

The pulses seen from PSR~B1957+20 have sub-microsecond timescales and are several
times stronger than the mean pulse.  All four coincide with the main
emission component in a similar fashion to the giant pulses of
PSR~J1823$-$3021A.  Even without evidence for power-law statistics it
is tempting to categorize PSR~B1957+20 as a giant pulse emitter.
However, the pulses can also be explained as strong pulses of ordinary
emission that are exceptionally narrow.  This hypothesis is supported
by the fact that pulses from PSR~J0437$-$4715 exhibit an
anti-correlation between pulse width and pulse strength
\citep{jak+98}.  The strongest ``ordinary'' pulses from
PSR~J0437$-$4715 are then much more readily detected in single-pulse
searches, and therefore could potentially masquerade as a giant-like
population.  Jenet \etal found pulses as short as 10\,$\mu$s in their
$\sim 3000$\,s of observations.  Therefore it is not unreasonable to
suggest that in $\sim 8000$\,s PSR~B1957+20 could emit several
``ordinary'' pulses consisting of very short spikes superimposed on
microsecond-timescale emission bursts.  The fact that we did not
detect any microsecond-timescale emission from PSR~J1012+5307 means
the pulse substructure seen by \citet{es03a} is broader than that seen
for PSR~B1957+20.  Similarly, \citet{kbmo05} did not find any
substructure in pulses from PSR~J1603$-$7202 as short as their
4\,$\mu$s sampling time.  Microstructure within ordinary pulses from
\msps therefore does not seem to have characteristic timescales as
short as those of the PSR~B1957+20 pulses.

Insight into whether or not the pulses from PSR~B1957+20 are plausibly
``giant'' can be gained by comparing the properties of PSR~B1957+20
and pulsars that emit giant pulses.  Table \ref{tab:characteristics}
summarizes the attributes of the \msps previously known to emit giant
pulses (top) and the pulsars we observed (bottom).  Each of these two
groups is sorted by right ascension.  The first three columns show the
pulsar name, period, and period derivative respectively.  The period
derivatives have been corrected for kinematic effects where possible
\citep{shk70,dt91}.  PSRs B1821$-$24 and J1823$-$3021A are located
within globular clusters, and so acceleration in the cluster potential
will contribute to the observed \pdot for these pulsars.  The
magnitude of the cluster contribution to \pdot is very uncertain, but
has been estimated to be $\le 0.06$\pdot for PSR~B1821$-$24
\citep{phi93} and $\le 0.7$\pdot for PSR~J1823$-$3021A \citep{sta97}.
The proper motions of PSRs J0218+4232 and J1843$-$1113 are unknown,
but a 100\,km\,s$^{-1}$ velocity equates to a Shklovskii-term
contribution to \pdot of just 0.6\% for PSR~J0218+4232 and 10\% for
PSR~J1843$-$1113.  Columns 4-7 of Table \ref{tab:characteristics} show
derived quantities -- the characteristic age, the magnetic field at
the light cylinder, the spin-down luminosity, and the complexity
parameter ($a_{\rm c} \approx 5 (\dot{P}/10^{-15})^{2/7} P^{-9/14}$)
as presented by \citet{gs00}.  Column 8 gives spectral indices
($\alpha_{\rm spec}$) and column 9 summarizes the X-ray luminosities
of the pulsars.  These are given for the 2-10 keV band unless
otherwise stated.

PSR~B1957+20 has comparable values of \blcs, $\dot{E}$, and $a_{\rm
c}$ to the four \msps that emit giant pulses.  Young pulsars like
PSR~B0540$-$69 and the Crab have much higher values of $\dot{E}$ and
$a_{\rm c}$, but they also emit many more giant pulses.  If any of
these attributes dictate \gp emissivity, we would expect PSR~B1957+20
to emit giant pulses.  If the pulses we see are not giant pulses, then
it is plausible that there is another population of pulses that has an
even lower rate of emission.  Presumably these pulses would take the
form of very narrow spikes that are restricted in pulse phase, just
like the pulses we see.  As invoking two populations of identical
looking pulses is contrived, we believe that we have seen purely giant
pulse emission, or giant pulse emission superimposed on a base of
ordinary emission.  An alternate idea that we do not favor is that at
moderate energies ordinary and giant pulses are indistinguishable
because the two seemingly disparate populations share a common
emission mechanism.  The ordinary pulses of PSR~B1937+21 show no sign
of modulation \citep{jg04} and the giant pulses only marginally
coincide with the envelope of ordinary emission.  The emission
mechanisms are therefore quite distinct for PSR~B1937+21, and
consequently it seems unlikely that the population of pulses from
PSR~B1957+20 represents the transition of a single pulse-population
from ordinary-like to giant-like emission.  At this stage we cannot
definitely state that PSR~B1957+20 emits giant pulses.  Further
supporting evidence could be made by establishing power-law statistics
and finding a correlation in phase with an X-ray pulse.  The general
task of identification of weak pulses as giant pulses is more
difficult.  Evidence could include giant pulses having different DMs
to ordinary pulses, or characteristic timescales much shorter than
those seen for microstructure.

Although PSR B1957+20 has a similar \blc to the \msps that emit giant
pulses, its emission rate is significantly lower.  In particular, its
rate would appear to be $\sim 100$ times lower than PSR~B1823$-$3021A,
despite the fact that PSR~B1957+20 has a higher $B_{\rm LC}$.
Magnetic inclination angle and other geometric factors must play some
role, but it is difficult to see how they could account for such an
enormous difference in emissivity.  So although the magnetic field at
the light cylinder does seem to be a reasonable determinant of whether
or not a pulsar emits giant pulses, it alone is not a trustworthy
indicator of the rate of emissivity.

\subsection{Giant Pulse Emitters}\label{sec:gpemitters}

PSR~J0218+4232 is the fourth \msp found that has been shown
conclusively to emit giant pulses.  All four such \msps have high
values of $\dot{E}$ and the complexity parameter.  The three observed
in X-rays are very luminous in the 2-10\,keV band and have hard photon
indices (see Table \ref{tab:characteristics} and references therein).
It is tempting to suggest that one or more of these characteristics
are better indicators of emissivity rates than \blcs.  However, PSRs
B1957+20 and J1843$-$1113 do not have corresponding values that are so
much lower that $\dot{E}$ and the complexity parameter can be
discriminated from \blc as the primary determinant of whether or not a
\msp emits giant pulses.  In fact \blcs, $\dot{E}$, and the complexity
parameter have such similar $P$-$\dot{P}$ dependences that we do not
think observations of \msps can ever discriminate between them.

Which other \msps could emit giant pulses?  Pulsars with high \blc
still seem to be good candidates, but this parameter no longer appears
to guarantee a rate sufficiently large to give a high detection count.
Table \ref{tab:characteristics} shows that the \msps that emit giant
pulses all have spectral indices much steeper than the average
$\alpha_{\rm spec}=-1.9$ spectrum \msps found by \citet{tbms98}.  They
also have very low characteristic ages and high X-ray luminosities.
Perhaps better sources are young or X-ray luminous pulsars in globular
clusters?  Unfortunately the Galactic globular cluster population is
old and so most cluster pulsars are likely to be too old to be good
candidates for \gp emission.  Consider PSR~J0024$-$7204J, which has a
0.5-6\,keV X-ray flux of $L_{\rm X} = 2 \times
10^{31}$\,ergs\,s$^{-1}$ \citep{hge+05}.  This is a similar luminosity
to PSR~B1957+20, so we do not expect PSR~J0024$-$7204J to emit giant
pulses at a high rate.  Since PSR~J0024$-$7204J has the highest X-ray
luminosity of the identified \msps in 47 Tucanae, we do not consider
47 Tucanae to be a good candidate cluster for giant pulse emission.
The clusters most likely to host populations of the young and X-ray
luminous millisecond pulsars prone to emitting giant pulses are
instead those that appear to contain young pulsars, such as the
core-collapsed clusters M15 and NGC 6624.

Perhaps all \msps emit giant pulses at even lower rates than
PSR~B1957+20?  The best candidates for verifying this hypothesis are
nearby \msps that have pulses that are not significantly
scatter-broadened.  Should bright pulsars like PSR~J0437$-$4715 emit
nanosecond-timescale pulses, then high time-resolution studies could
potentially probe their pulse populations down to very low energies.
Such studies could reveal giant pulses occuring at rates smaller by
factors of $\sim 1000$ than seen for PSR~B1957+20.

\section{Conclusions}\label{conclusions}

We have searched four millisecond pulsars for individual pulses of
emission with microsecond timescales and have found such emission from
two of them.  Only four individual pulses were detected from
PSR~B1957+20 in 8003\,s of observations centered at 825\,MHz.  As
these pulses are exceptionally narrow there is little
scattering-induced pulse-broadening at least some orbital phases.
Although it is debatable whether or not these strong pulses are true
``giant pulses'', we can say that the giant pulse emission rate from
PSR~B1957+20 is significantly less than the rates for other pulsars
with similar values of magnetic field at the light cylinder.  Although
\blc can be used as a rough guide to whether a pulsar emits giant
pulses, we suggest it is a poor indicator of the emission rate.

PSR~J0218+4232 emits giant pulses at a low rate that is inconsistent
with the findings of \citet{jkl+04a}.  It is most likely that the pulses
reported by Joshi \etal are spurious.  The giant pulses of PSR~J0218+4232 are
confined to two narrow phase regions separated by roughly 50\% of
phase which align in phase with the peaks of the X-ray profile and
roughly coincide with the minima of the integrated pulse-profile in
the radio band.  This strong correlation between X-ray and radio
properties confirms that the two emission processes originate in
similarly defined regions of the pulsar magnetosphere.

Most of the 139 giant pulses observed from PSR~J0218+4232 at a center
frequency of 857\,MHz had relatively low energies, typically only a few
times the mean pulse energy.  Only three had energies above $10\eav$
and none had energies above $20\eav$.  The pulses exhibit power-law
statistics, are only found in narrow phase windows that coincide in
phase with the X-ray pulse-components, and are very narrow just like
the giant pulses of PSR~B1937+21; it is apparent then that ``giant''
pulses should be defined not through large flux densities, but by
these three properties.  The brightest pulse seen at a center
frequency of 1373\,MHz seems to be around 500\,ns in duration when
viewed at 125\,ns time resolution.  At higher time resolution finer
features become apparent, but it is unclear whether these are
significant.

PSR~J0218+4232 is the fourth \msp found to emit giant pulses after
PSRs B1937+21, B1821$-$24, and J1823$-$3021A.  All four have low
characteristic ages and steep radio spectra.  With the exception of
PSR~J1823$-$3021A which has not been observed in X-rays, the four
pulsars all have high X-ray luminosities and exhibit power-law
spectra.  The presence of X-ray emission with a steep power-law
spectrum therefore seems to be the best indicator of whether a \msp
emits giant pulses.  Radio observations would be expected to show that
narrow giants will be present at the phase of the X-ray emission.

\acknowledgements

The National Radio Astronomy Observatory is a facility of the National
Science Foundation operated under cooperative agreement by Associated
Universities, Inc.  We thank L. Kuiper for providing high energy data.
HSK acknowledges the support of a CSIRO Postgraduate Student Research
Scholarship.  BAJ thanks NSF and NASA for supporting this research.

\clearpage

\bibliographystyle{apj}

\clearpage

\input{tab1.tex}

\clearpage

\input{tab2.tex}

\clearpage

\input{tab3.tex}

\clearpage

\clearpage

\clearpage

\clearpage

\end{document}

%% file: tab1.tex
\begin{deluxetable}{lrccccccc}
\tablewidth{0pt}
\tablecaption{\label{tab:observations}Summary of searches for \gp emission.}
\tablecolumns{9}
\tablehead{ 
\colhead{PSR} & 
\colhead{$\nu$} & 
\colhead{$\delta \nu$} & 
\colhead{$t_{\rm obs}$} & 
\colhead{$N_{\rm p}$} & 
\colhead{$\eav$} & 
\colhead{$E_{\rm lim}$} & 
\colhead{$N_{\rm det}$} & 
\colhead{Notes} \\ 
 &
\colhead{(MHz)} & 
\colhead{(MHz)} & 
\colhead{(s)} & 
\colhead{($\times 10^5$)} & 
\colhead{(${\rm Jy}\cdot{\rm \mu s}$)} & 
\colhead{($\eav$)} & 
\colhead{} & 
\colhead{} \\ 
}

\startdata 
J0218+4232  & 857  & 128 & 3456    & 15     & 18     & 0.58      & 24, 75   & (1) \\
            & 857  & 128 & 5216    & 22     & 9.2    & 0.95      & 5, 35   & (2) \\
            & 1373 & 64  & 3349    & 14     & 3.9    & 2.5       & 4, 8   & \\
            & 825  & 64  & 293     & 1.3    & 8.7    & 1.7       & 0, 4    & \\
J1012+5307  & 825  & 64  & 720     & 1.4    & 60     & 0.27      & 0    & \\
            & 1373 & 64  & 6077    & 12     & 15     & 0.72      & 0    & \\
            & 1437 & 64  & 6951    & 13     & 22     & 0.51      & 0    & \\
J1843$-$1113& 825  & 64  & 3198    & 17     & 6.2    & 4.3       & 0    & \\
            & 1373 & 64  & 791     & 4.3    & 2.3    & 5.8       & 0    & \\
            & 1437 & 64  & 1436    & 7.8    & 1.9    & 6.9       & 0    & \\
J1959+2048  & 825  & 64  & 8003    & 50     & 4.7    & 4.0       & 4    & \\
            & 1373 & 64  & 4684    & 29     & 0.72   & 16        & 0    & (3) \\
            & 1437 & 64  & 4760    & 30     & 0.63   & 19        & 0    & (3) \\
\enddata

\tablecomments{
(1) 2004 August observation; (2) 2004 October observation; (3) No phase-coherent timing solution was available because of incorrect time-tagging.  The flux density ($S$) used to derive the given parameters is given by the $S = 0.35 (\nu/$1490 MHz$)^{-3}$\,mJy relation of \citet{fbb+90}.
}

\end{deluxetable}

%% file: tab2.tex
\begin{deluxetable}{lrccccccc}
\tabletypesize{\tiny}
\tablewidth{0pt}
\tablecaption{\label{tab:powerlawfits}Giant pulse emission rates from a selection of pulsars.}
\tablecolumns{9}

\tablehead{

\colhead{PSR} & 
\colhead{$\nu$ (MHz)} & 
\colhead{Phase range} & 
\colhead{$K$} & 
\colhead{$\alpha$} & 
\colhead{$P(E > 20\eav)$} & 
\colhead{$S_{\rm GP}(E > 20\eav)$} & 
\colhead{$S_{\rm GP}(E > 0.1\eav)$} & 
\colhead{References} \\ 
}

\startdata

J0218+4232&857 &A (August) &$1.2\times 10^{-5}$&1.5&$1.3\times 10^{-7}$&$8.0\times 10^{-6}$&$1.1\times 10^{-4}$& This work.\\
J0218+4232&857 &B (August) &$5.6\times 10^{-5}$&1.9&$1.9\times 10^{-7}$&$8.0\times 10^{-6}$&$9.4\times 10^{-4}$& This work.\\
J0218+4232&857 &B (October)&$4.4\times 10^{-5}$&1.7&$2.7\times 10^{-7}$&$1.3\times 10^{-5}$&$5.4\times 10^{-4}$& This work.\\
J0218+4232&1373&all        &$3.4\times 10^{-5}$&1.5&$3.8\times 10^{-7}$&$2.3\times 10^{-5}$&$3.2\times 10^{-4}$& This work.\\
B1957+20\tablenotemark{a}&825 &main pulse &$2\times 10^{-5}$  &2  &$5\times 10^{-8}$  &$2\times 10^{-6 }$ &$4\times 10^{-4 }$ & This work.\\

B0531+21 (Crab)& 146  & main pulse &$2.8\times 10^{-1}$& 2.5 &$1.5\times 10^{-4}$&$5.1\times 10^{-3}$& n/a        & (1)\\
B0531+21 (Crab)& 800  & all        &$9.8              $& 2.4 &$8.3\times 10^{-3}$&$2.9\times 10^{-1}$& n/a        & (2)\\
B0540$-$69     & 1390 & early      &$2.4\times 10^{-2}$& 1.5 &$2.7\times 10^{-4}$&$1.6\times 10^{-2}$& n/a        & (3)\\
B0540$-$69     & 1390 & late       &$7.6\times 10^{-1}$& 2.1 &$1.4\times 10^{-3}$&$5.3\times 10^{-2}$& n/a        & (3)\\
B0540$-$69     & 1390 & all        &$2.6\times 10^{-1}$& 1.8 &$1.2\times 10^{-3}$&$5.4\times 10^{-2}$& n/a        & (3)\\
B1937+21       & 430  & all        &$3.2\times 10^{-2}$& 1.8 &$1.5\times 10^{-4}$&$6.6\times 10^{-3}$& n/a        & (4)\\
B1937+21       & 1650 & all        &$2.8\times 10^{-4}$& 1.4 &$4.2\times 10^{-6}$&$3.0\times 10^{-4}$&$2.5\times 10^{-3}$&(5)\\
\enddata

\tablenotetext{a}{ Rates are indicative only due to the very small number of pulses analysed.}

\tablerefs{
(1) \citet{ag72}; (2) \citet{lcu+95}; (3) \citet{jrmz04}; (4) \citet{cstt96}; (5) \citet{spb+04}.
}

\end{deluxetable}

%% file: tab3.tex
\begin{deluxetable}{lrcccccclc}
\tabletypesize{\tiny}
\tablewidth{0pt}
\tablecaption{\label{tab:characteristics}Pulsar characteristics.}
\tablecolumns{10}

\tablehead{ 

\colhead{PSR} & 
\colhead{$P$} & 
\colhead{\pdot} & 
\colhead{$\tau$} & 
\colhead{\blc} & 
\colhead{$\dot{E}$} & 
\colhead{$a_{\rm c}$} & 
\colhead{$\alpha_{\rm spec}$} & 
\colhead{$L_{\rm X (2-10\,keV)}$} & 
\colhead{References} \\ 
 &
\colhead{(ms)} & 
\colhead{($10^{-21}$)} & 
\colhead{(Myr)} & 
\colhead{($10^4$G)} & 
\colhead{($10^{33}$ ergs s$^{-1}$)} & 
\colhead{} & 
\colhead{} & 
\colhead{($10^{32}$ ergs s$^{-1}$)} & 
\colhead{} \\ 

}

\startdata

J1823$-$3021A\tablenotemark{a}&5.44& 3390  & 26   & 25  & 810  & 28 & -2.7 & Unknown                         & (1,2)\\
B1821$-$24\tablenotemark{a}&3.05  & 1620  & 30   & 74  & 2200 & 33 & -2.3 & 13, 12.8 (0.5-8\,keV)           & (1,3,4,5)\\
B1937+21                   & 1.56  & 106   & 230  & 102 & 1100 & 23 & -2.6 & 0.5-5.7                         & (1,3,6,7,8)\\
\hline
J0218+4232                 & 2.32  & 77.4  & 480  & 32  & 240  & 16 & -3.0 & 1.3 (1-10\,keV); 1.2-1.6        & (1,9,10,11,12)\\
J1012+5307                 & 5.26  & 9.73  & 8600 & 1.5 & 2.6  & 6.3& -1.9 & 0.003 (0.2-10\,keV)             & (13,14,15)\\
J1843$-$1113               & 1.85  & 9.59  & 3100 & 20  & 59   & 11 & Unknown & Unknown                      & (1)\\
B1957+20                   & 1.61  & 11.5  & 2200 & 31  & 110  & 14 & -3.0 & 0.16 (0.5-7\,keV)               & (1,6,16,17)\\

\enddata

\tablenotetext{a}{ Parameters ignore acceleration in the gravitational potential of the host cluster.}

\tablerefs{
(1) \citet{mhth05}; (2) \citet{tbms98};
(3) \citet{ffb91}; (4) \citet{bsp+03}; (5) \citet{mcm+04};
(6) \citet{tsb+99}; (7) \citet{tst+01}; (8) \citet{ncl+04};
(9) \citet{nbf+95}; (10) \citet{mck+00}; (11) \citet{khv+02}; (12) \citet{wob04};
(13) \citet{nll+95}; (14) \citet{lcw+01}; (15) \citet{wob+04};
(16) \citet{fbb+90}; (17) \citet{sgk+03}.}

\end{deluxetable}